# ANALYSIS AND CONTROL OF PERIOD DOUBLING BIFURCATION IN BUCK CONVERTERS USING HARMONIC BALANCE

C.-C. FANG† and E. H. ABED‡

†*Logic Library Dept., Taiwan Semiconductor Manufacturing Co., Hsinchu 300, Taiwan*
*ccfang@isr.umd.edu*
‡*Department of Electrical and Computer Engineering and the Institute for Systems Research*
*University of Maryland, College Park, MD 20742, USA*
*abed@isr.umd.edu*

***Abstract*— Period doubling bifurcation in buck converters is studied by using the harmonic balance method. A simple dynamic model of a buck converter in continuous conduction mode under voltage mode or current mode control is derived. This model consists of the feedback connection of a linear system and a nonlinear one. An exact harmonic balance analysis is used to obtain a necessary and sufficient condition for a period doubling bifurcation to occur. If such a bifurcation occurs, the analysis also provides information on its exact location. Using the condition for bifurcation, a feedforward control is designed to eliminate the period doubling bifurcation. This results in a wider range of allowed source voltage, and also in improved line regulation.**

## I. INTRODUCTION

Several authors have investigated the occurrence of period doubling bifurcation in DC-DC converters (Deane and Hamill, 1990; Hamill *et al.*, 1992; Tse, 1994; Fossas and Olivar, 1996). A period doubling bifurcation entails loss of stability of the nominal operating condition, and as such is undesirable. Moreover, a period doubling route to chaos could be signaled by such a bifurcation, eroding the performance of the circuit. This has led to preliminary investigations of methods for prevention of period doubling bifurcations in DC-DC converters for a *single* value of source voltage in (Podder *et al.*, 1995; Poddar *et al.*, 1998).

In this work, analysis and control of period doubling bifurcation of a buck converter in continuous-conduction mode are considered. A continuous-time feedback system model is used as the basis for a harmonic balance analysis of period doubling bifurcation as well as control design for preventing the onset of period doubling. The model separates the nonlinear switching action of the converter from the linear filtering action. Because the model resolves dynamics within switching intervals, it is more accurate than the traditional averaged models (Middlebrook and Ćuk, 1976).

In addition, the model proposed here is valid both for current mode control and voltage mode control, and can be applied to *larger range* of source voltage.

An *exact* harmonic balance analysis is used to obtain a necessary and sufficient condition for a period doubling bifurcation to occur. If such a bifurcation occurs, the analysis also provides information on its exact location. Using the condition for bifurcation, a feedforward control is designed to eliminate the period doubling bifurcation. This results in a larger range of allowed source voltage, and also in improved line regulation. (Line regulation entails regulating the output voltage close to a constant for different values of source voltage.)

Harmonic balance analysis of period doubling bifurcations has been pursued for general nonlinear systems in (Genesio and Tesi, 1992; Piccardi, 1994; Tesi *et al.*, 1996). Unlike the approximate but general treatments given in these references, in this paper an *exact* harmonic balance analysis is performed. This is made possible by the special structure of the buck converter model. For other converters, a more typical approximate harmonic balance analysis can be performed.

It is noteworthy that first-order harmonic balance analysis has previously been applied in power electronics in the context of small-signal modeling (Yang, 1994; Groves Jr., 1995).

The remainder of the paper is organized as follows. In Section II, a simple dynamic model is proposed for the buck converter in continuous conduction mode under voltage mode or current mode control. In Section III, the harmonic balance method is used to study period doubling bifurcation for the buck converter. In Section IV, feedforward control is used to eliminate period doubling bifurcation and improve line regulation. In Section V, an illustrative example is given.

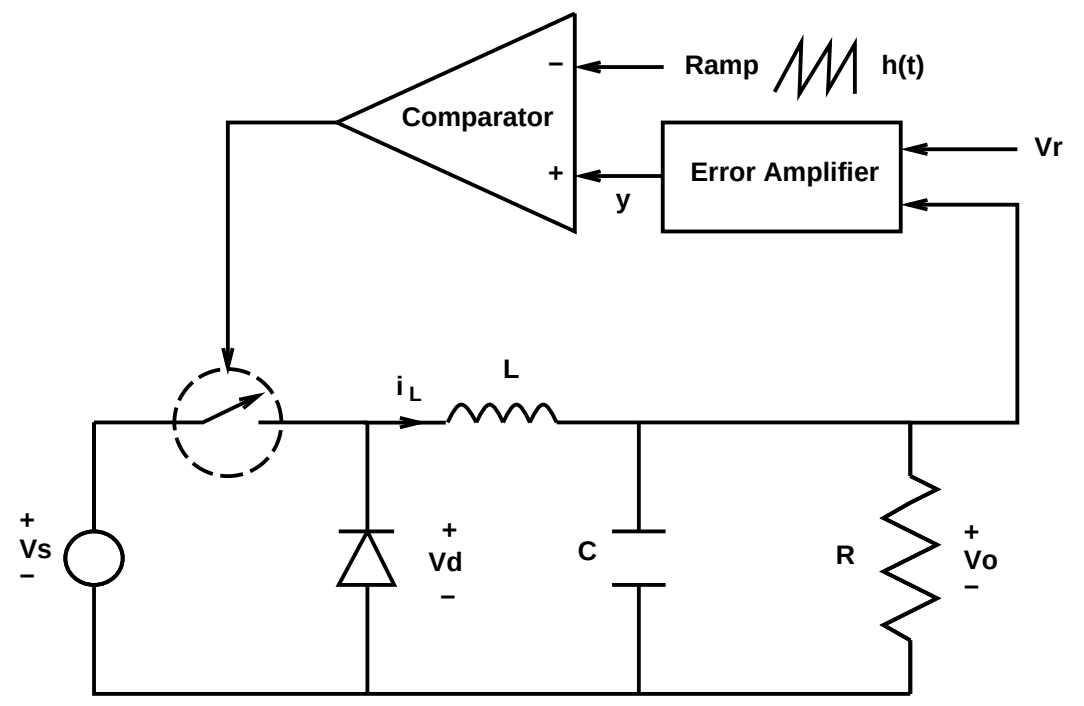

Figure 1: Buck converter under voltage mode control

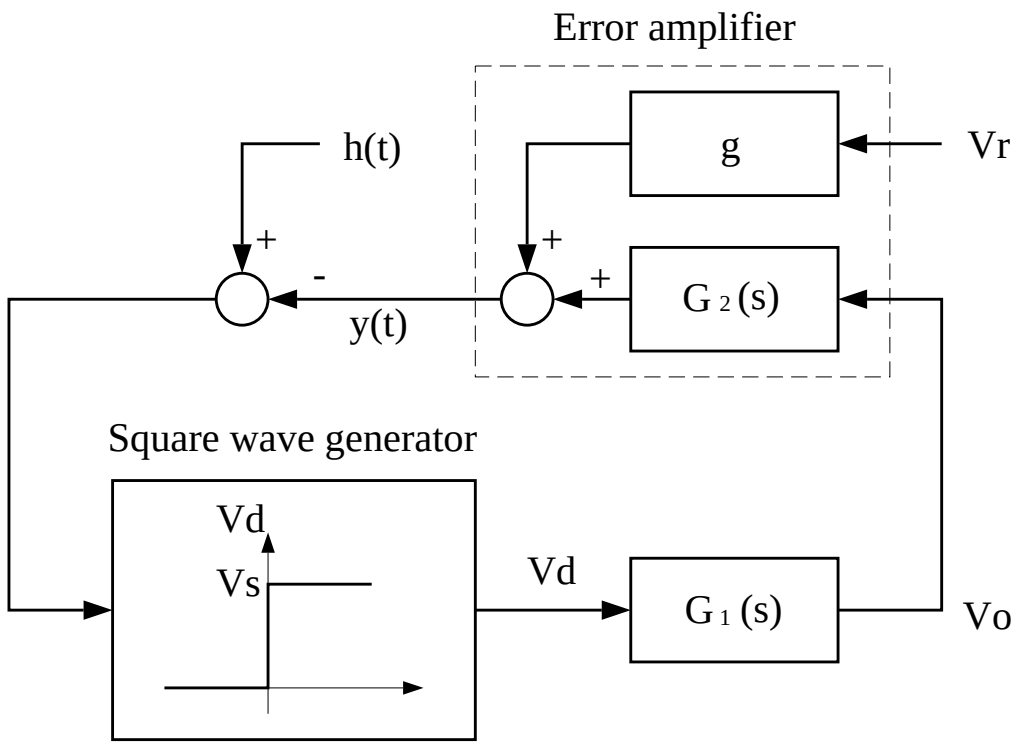

Figure 2: Dynamic model for voltage mode control

Conclusions are collected in Section VI.

## II. SIMPLE DYNAMIC MODEL

A simple dynamic model is proposed for the buck converter in continuous conduction mode under voltage mode control or current mode control.

### A. Voltage Mode Control

A buck converter under voltage mode control is shown in Fig. 1. The source voltage and reference voltage are assumed constant, and are denoted by $V_s$ and $V_r$, respectively. The output voltage is $v_o$. The voltage across the diode is denoted by $v_{\mathrm{d}}$. Output signal of the error amplifier is denoted by $y(t)$. The ramp signal, denoted by $h(t) = V_l + (V_h - V_l)(\frac{t}{T} \bmod 1)$, is a $T$-periodic function (with $h(0) = V_l$ and $h(T) = V_h$). The switching period is $T$, the switching frequency is $f_s = 1/T$, and the angular switching frequency is $\omega_s = 2\pi f_s$.

The switching operation in continuous conduction mode (with leading-edge modulation) is as follows. When $y(t) < h(t)$, the switch is on, the diode is off, and $v_{\mathrm{d}} = V_s$. When $y(t) \geq h(t)$, the switch is off, the diode is on, and $v_{\mathrm{d}} = 0$. Thus the signal $v_{\mathrm{d}}(t)$ is a square wave, and the switch/diode combination can be modeled as a controlled square wave generator. In this paper, leading-edge modulation is assumed. Similar analysis can be readily applied to the case of trailing-edge modulation, where the switch is on for $y(t) \geq h(t)$ and the switch is off for $y(t) < h(t)$.

The output filter ($L$ and $C$ with equivalent series resistance (ESR) $R_c$) and the load ($R$) in Fig. 1 form a low-pass filter with transfer function

$$G_1(s) = \frac{R_c C s + 1}{LC(1 + \frac{R_c}{R})s^2 + (\frac{L}{R} + R_c C)s + 1} \qquad (1)$$

For a buck converter with a more complex power stage (for example, with a second output filter), the model remains valid, but with a more complex transfer function $G_1(s)$.

Generally the error amplifier is linear and is driven by the signals $V_r$ and $v_o$. Its output can be represented as

$$y(t) = gV_r + (g_2 \star v_o)(t) \qquad (2)$$

where $g$ is a gain constant, $g_2(t)$ is an impulse response function and $\star$ denotes convolution. Denote the transfer function associated with $g_2(t)$ by $G_2(s)$. Since $G_2(s)$ depends on the control scheme, no further restriction is placed on it.

Thus the buck converter in continuous conduction mode can be modeled as a system block diagram shown in Fig. 2. The output of the controlled square wave generator is $V_s$ if $(h(t) - y(t))$ is greater than 0, the output is 0 if $(h(t) - y(t))$ is less than 0.

### B. Current Mode Control

For a buck converter under current mode control as depicted in Fig. 3, the switch operation is different from that in voltage mode control. In current mode control, the switch turns on at each clock pulse and turns off at instants when $y(t) = h(t)$. Here $h(t)$ is a slope-compensating ramp. The switch/diode combination can also be modeled as a controlled square wave generator. Therefore, a buck converter under current mode control can be modeled by the system block diagram in Fig. 4, where a linear transfer function

$$\begin{aligned} G_i(s) &= \frac{R_s i_L(s)}{v_{\mathrm{d}}(s)} \\ &= \frac{R_s(RC(R+R_c)s+1)}{RLC(R+R_c)s^2 + (L+RR_cC)s + R} \end{aligned}$$

is placed in the current feedback path. The remaining notation is the same as for the case of voltage mode control.

### C. Unified Dynamic Model

Despite the differences in switching operation in voltage mode control and current mode control, they can be modeled in a unified setting. Both Fig. 2 and Fig. 4 can be further simplified as Fig. 5, where $G(s) = G_1(s)G_2(s)$ for voltage mode control and $G(s) = G_1(s)$ $G_2(s) - G_i(s)$ for current mode control. For either case, the controlled buck converter is a combination of a linear system $G(s)$ and a nonlinear square wave generator.

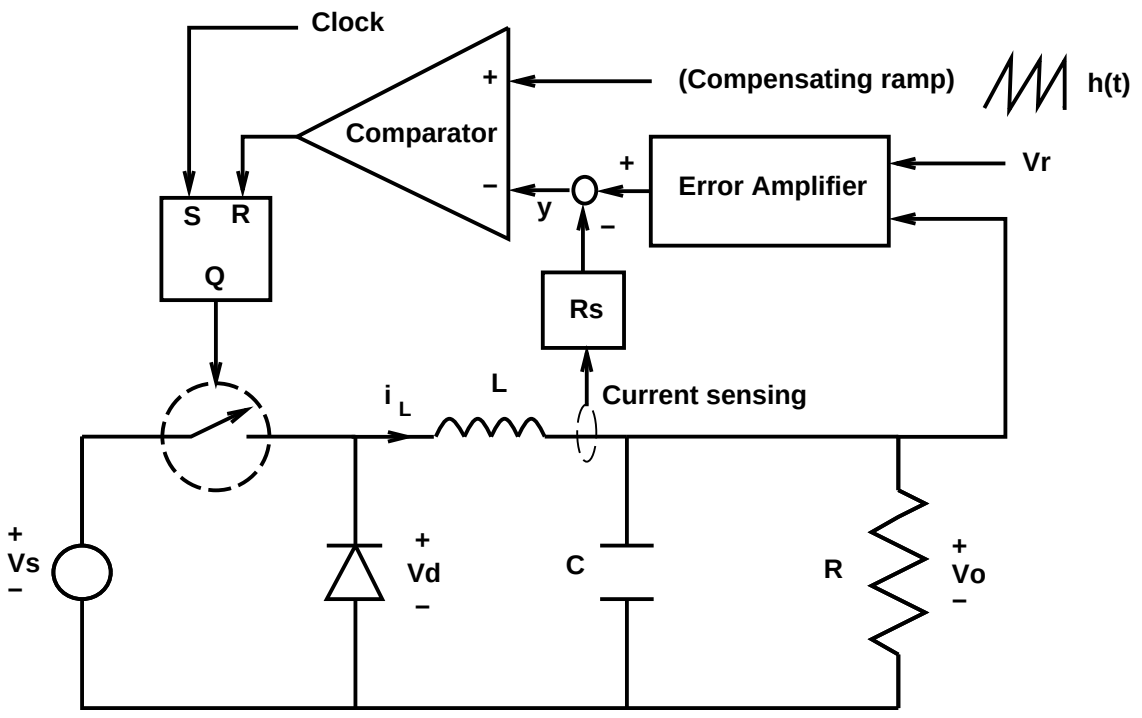

Figure 3: Buck converter under current mode control

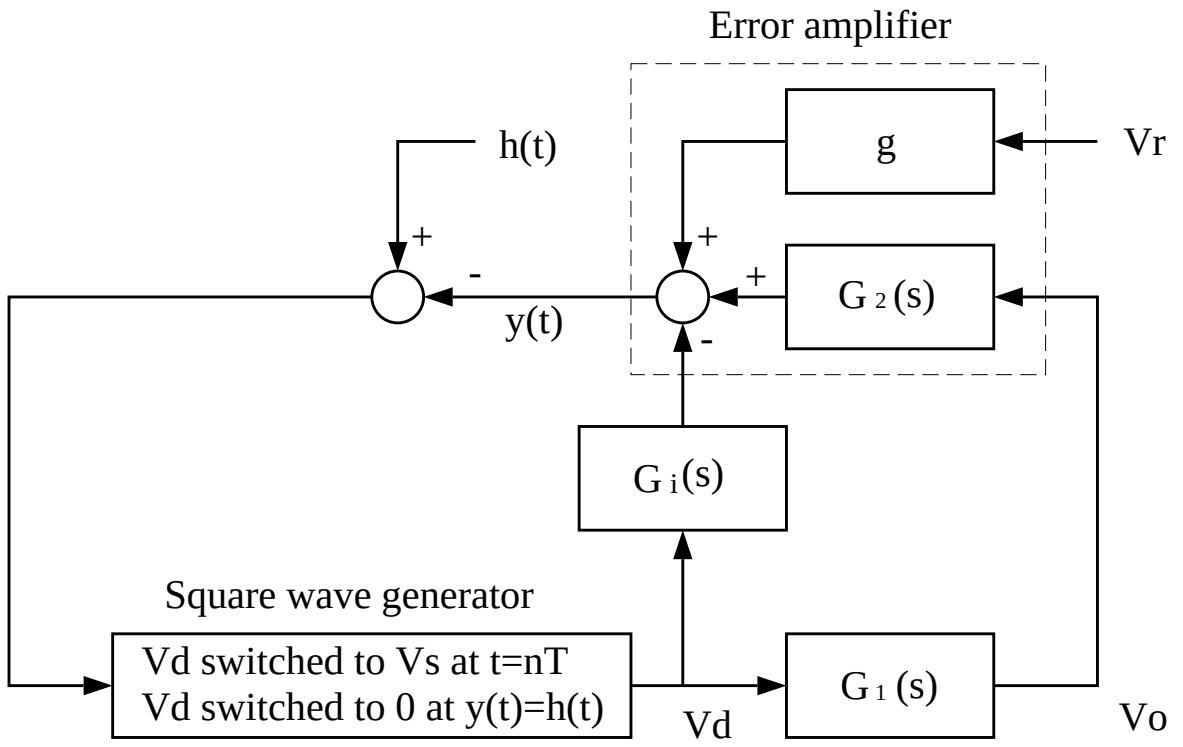

Figure 4: Dynamic model for current mode control

In the remainder of the paper, more emphasis is put on voltage mode control. The results extend readily to the case of current mode control.

## III. DETERMINATION OF PERIOD DOUBLING BIFURCATION POINT

In practice, the nominal operating condition is a $T$-periodic solution (in period-one mode). Representative waveforms for $y(t)$ and $v_{\mathrm{d}}(t)$ are shown in Fig. 6. At the switching instant $t = d$,

$$y(d) = h(d) \tag{3}$$

When period doubling bifurcation occurs, a $2T$-periodic solution arises from the original $T$-periodic solution. Representative $2T$-periodic waveforms for $y(t)$ and $v_{\mathrm{d}}(t)$ are depicted in Fig. 7. Switchings occur at $t = d - \delta$ and at $T + d + \delta$, where $\delta$ is a small parameter that vanishes at the bifurcation point. From the switching conditions at these two instants, it follows that

$$\begin{aligned} y(d-\delta) &= h(d-\delta) & (4)\\ y(T+d+\delta) &= h(T+d+\delta) & (5) \end{aligned}$$

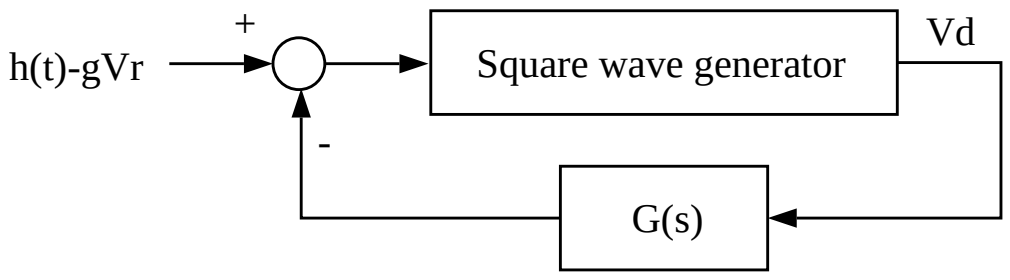

Figure 5: Unified dynamic model of buck converter

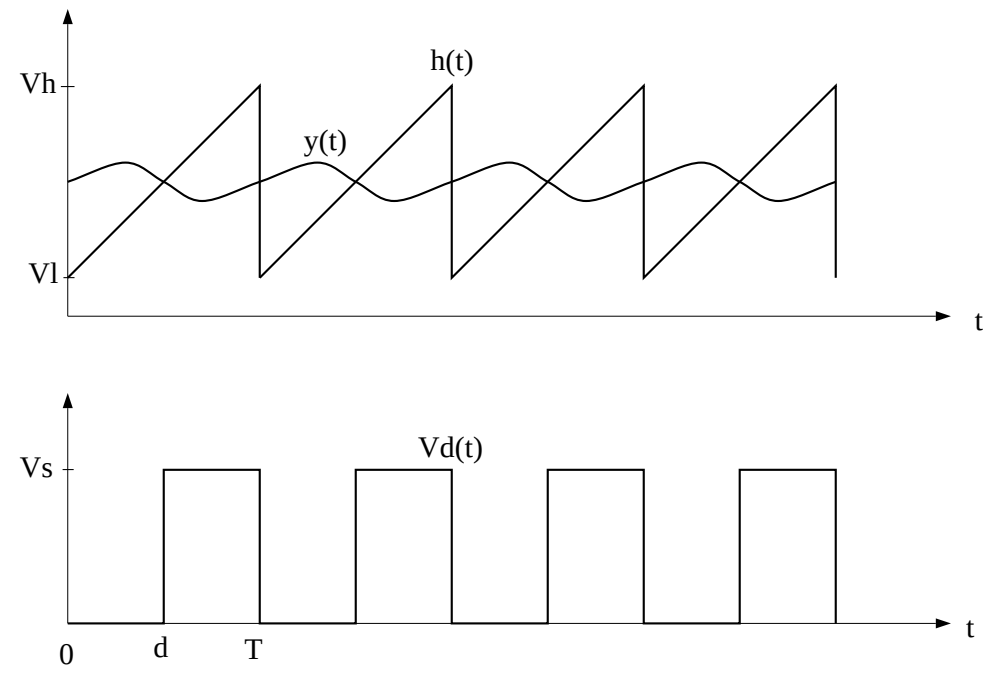

Figure 6: $y(t)$, $h(t)$ and $v_{\mathrm{d}}(t)$ in period-one mode

The harmonic balance method is a tool that can be used to analyze periodic solutions in nonlinear systems. In the buck converter, the nonlinearity results from the switch. In the steady-state, $v_{\mathrm{d}}$ is a periodic signal and can be represented by a Fourier series. By "balancing" the equations above (written in Fourier series form) at the switching instants, a condition for existence of a periodic solution is derived.

The harmonic balance method is applied here to determine conditions for a period doubling bifurcation to occur. The basic idea is that at such a bifurcation point (shown in Fig. 8), a period-one mode and a period-two mode coalesce. By invoking conditions for the existence of each mode, the period doubling bifurcation point can be determined. In the following, the source voltage $V_s$ is used as the bifurcation parameter. The critical value of the bifurcation parameter, denoted by $V_{s,*}$, is determined. If a different parameter is taken as the bifurcation parameter, the approach is similar.

### A. Harmonic Balance of Period-One Mode

In the period-one mode, $v_{\mathrm{d}}(t)$ is periodic with angular frequency $\omega_s$, and can be expressed as the Fourier series

$$v_{\mathrm{d}}(t) = \sum_{n=-\infty}^{\infty} c_n e^{jn\omega_s t} \tag{6}$$

where

$$c_n = \frac{V_s}{j2n\pi}(e^{-jn\omega_s d} - e^{-jn\omega_s T})$$

Let the duty cycle be $D_c$. Then $D_c = 1 - d/T$. The average value of $v_{\mathrm{d}}(t)$, denoted as $[v_{\mathrm{d}}(t)]_{\mathrm{AVE}}$, is

$$[v_{\mathrm{d}}(t)]_{\mathrm{AVE}} = c_0 = (1 - d/T)V_s = D_c V_s \tag{7}$$

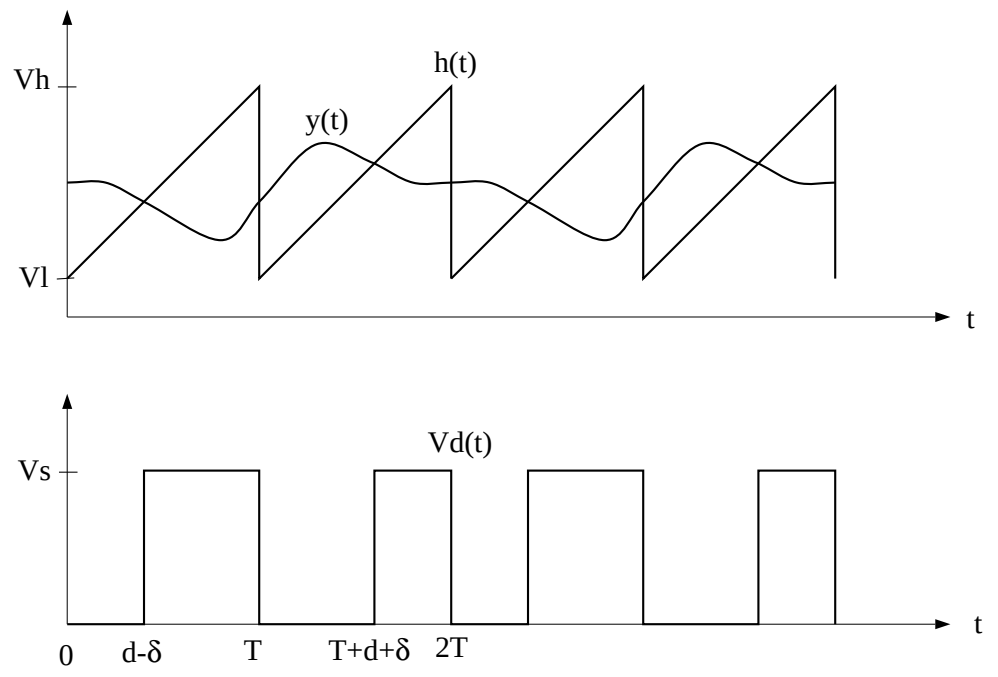

Figure 7: $y(t)$, $h(t)$ and $v_{\mathrm{d}}(t)$ in period-two mode

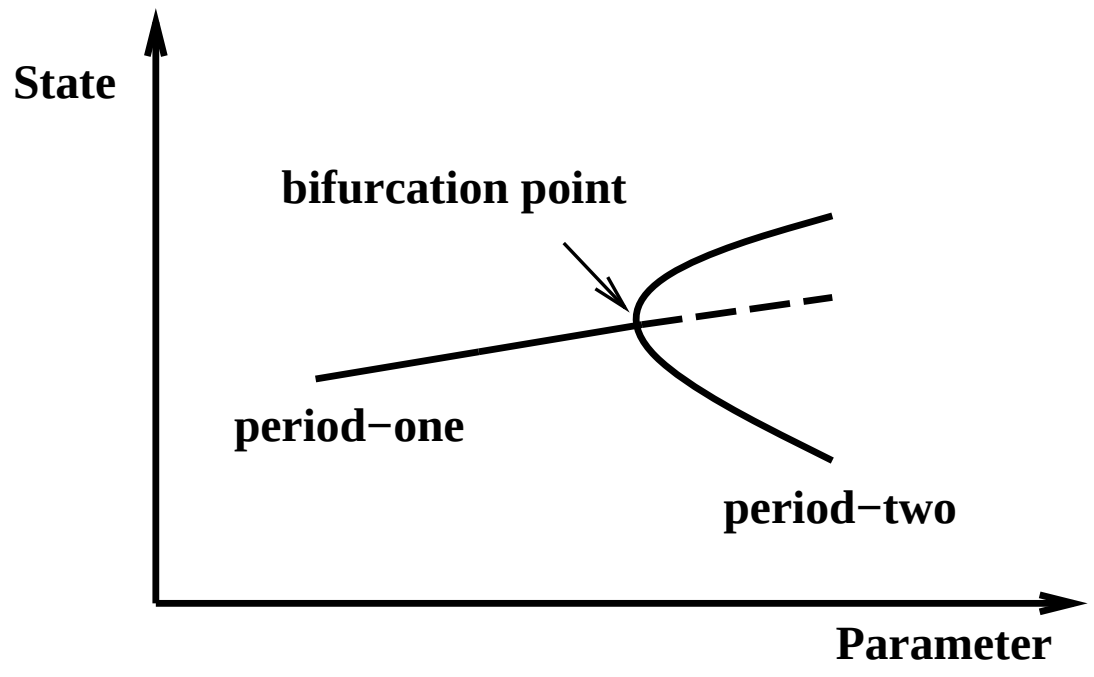

Figure 8: Period doubling bifurcation

This agrees with the standard result from the averaging method.

From Fig. 2, the signal at the output of the error amplifier is

$$\begin{aligned} y(t) &= gV_r + (g_2 \star v_o)(t) \\ &= gV_r + \sum_{n=-\infty}^{\infty} c_n e^{jn\omega_s t} G(jn\omega_s) \end{aligned} \tag{8}$$

Using Eqns. (8) and (3), $V_s$ can be written in terms of $d$ as

$$V_s = \frac{h(d) - gV_r}{(1 - \frac{d}{T})G(0) + (\frac{1}{\pi})\mathrm{Im}[\sum_{n=1}^{\infty} \frac{1-e^{jn\omega_s d}}{n} G(jn\omega_s)]} \tag{9}$$

## B. Harmonic Balance of Period-Two Mode

Similarly, in the period-two mode, $y(t)$ is $2T$-periodic and can be represented as the Fourier series

$$y(t) = gV_r + \sum_{n=-\infty}^{\infty} c_n e^{\frac{jn\omega_s t}{2}} G(\frac{jn\omega_s}{2}) \tag{10}$$

where

$$c_n = \begin{cases} (\frac{V_s}{n\pi}) e^{-\frac{jn\omega_s d}{2}} \sin(\frac{n\omega_s \delta}{2}), \text{if } n \text{ is odd} \\ (\frac{V_s}{jn\pi})(e^{-\frac{jn\omega_s d}{2}} \cos(\frac{n\omega_s \delta}{2}) - e^{-\frac{jn\omega_s T}{2}}), \text{if } n \text{ is even} \end{cases}$$

Subtracting (4) from (5) and substituting (10) for $y$ gives

$$\begin{aligned} &\delta \frac{V_h - V_l}{T} = \\ &\quad (\frac{V_s}{\pi})\mathrm{Re}(-\sum_{k=1}^{\infty} \frac{1}{2k-1} G(j(k-\frac{1}{2})\omega_s) \sin((2k-1)\omega_s \delta) \\ &\quad + \sum_{k=1}^{\infty} \frac{1}{2k} G(jk\omega_s)(\sin(2k\omega_s \delta) - 2e^{jk\omega_s d} \sin(k\omega_s \delta)) \end{aligned}$$

Solving for $V_s$ gives another expression for $V_s$ in terms of $d$ and $\delta$. See (Fang, 1997) for the detailed form.

## C. Determination of Bifurcation Point

Recall that $\delta$ is small and that, if a period doubling bifurcation occurs, then $\delta = 0$ at the bifurcation point. The critical value of $V_s$ at the bifurcation is then determined:

$$V_{s,*} = \frac{\frac{V_h - V_l}{2}}{\mathrm{Re}[-\sum_{k=1}^{\infty} G(j(k-\frac{1}{2})\omega_s) + (1 - e^{jk\omega_s d})G(jk\omega_s)]} \tag{11}$$

The critical values $V_{s,*}$ and $d_*$ can be obtained graphically by plotting Eqns. (9) and (11) on the same axes. The intersection $(V_{s,*},\ d_*)$ of these graphs (if it occurs) is the period doubling bifurcation point. Indeed, a necessary and sufficient condition for a period doubling bifurcation to occur is that these graphs intersect. (Period doubling bifurcation occurs when the source votage reaches $V_{s,*}$, or duty cycle reaches $d_*/T$.)

The denominator of (11) can be approximated by the term that involves $G$ with the smallest argument, namely, $\mathrm{Re}[G(j\omega_s) - G(\frac{j\omega_s}{2})]$. So a good estimate for the critical value $V_{s,*}$ is

$$\frac{\frac{V_h - V_l}{2}}{\mathrm{Re}[G(j\omega_s) - G(\frac{j\omega_s}{2})]} \tag{12}$$

Further bifurcation to the period-four mode can be analyzed similarly and is obmitted here. Two special cases are considered next.

**Example 1** Consider a buck converter under *voltage* mode control with $G_2(s) = g_1$ and $g = -g_1$, shown in Fig. 9. Under the conditions $1/\sqrt{LC} \ll \omega_s$, $R_c \ll R$, and $R_c C \ll 1$ (which are generally true), the critical source voltage (Eqn. (12)) can be further simplified as

$$V_{s,*} \approx (\frac{V_h - V_l}{6g_1})(\frac{R + R_c}{R}) LC\omega_s^2 \tag{13}$$

It shows that larger values of $\omega_s$, $L$, $C$, and ramp amplitude $(V_h - V_l)$ lead to larger stable range of source voltage, while larger feedback gain $g_1$ leads to smaller range of source voltage. Also ESR $R_c$ (compared to $R$) is generally small and has little effect on stability.

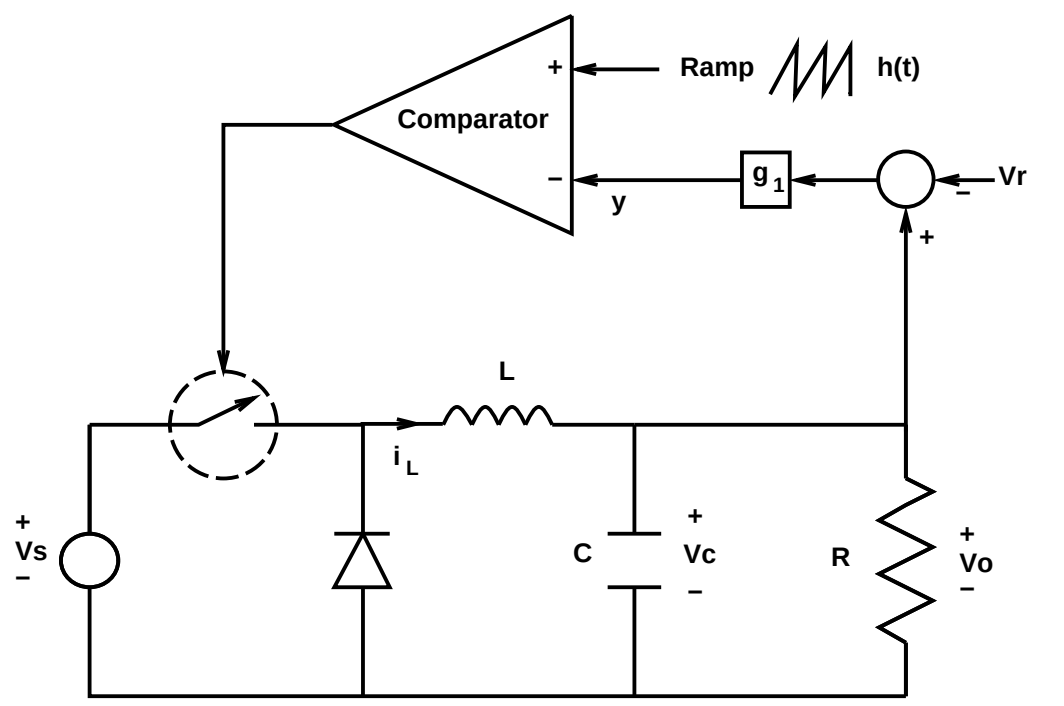

Figure 9: System diagram for Example 1

Critical values of other parameters can be obtained similarly from Eqn. (13). For example, the critical feedback gain $g_{1,*}$ is close to

$$(\frac{V_h - V_l}{6V_s})(\frac{R+R_c}{R})LC\omega_s^2 \tag{14}$$

**Example 2** Consider a buck converter under *current* mode control with open voltage loop ($G_2(s) = 0$). Under the conditions $\omega_s \gg (1/\sqrt{LC}$, $R_c/L$, and $1/RC)$, the critical source voltage (Eqn. (12)) can be further simplified as

$$V_{s,*} \approx (\frac{V_h - V_l}{6R_s})(\frac{R+R_c}{RR_c})L^2\omega_s^2 \tag{15}$$

With slope compensation and larger values of $\omega_s$ and $L$, the stable range of source voltage is larger. However, different from voltage mode control, larger ESR $R_c$ leads to smaller range of source voltage.

These two examples shows that if a control scheme is designed appropriately to adjust the amplitude of the ramp, the period doubling bifurcation can be prevented. This is pursued next.

## IV. FEEDFORWARD CONTROL

Common objectives of controller design for a DC-DC converter are stability, fast transient dynamics, line regulation, and load regulation. In some demanding situations, not all of these objectives can be achieved simultaneously due to nonlinear nature of the converter. In this section, feedforward control is used to prevent period doubling bifurcation (instability) and to achieve line regulation. Transient dynamics and load regulation can be improved by redesigning the voltage feedback loop, however, this is not the focus of this paper.

Here a feedforward control scheme is proposed to adjust the ramp signal $h(t)$ by setting $V_l = k_l V_s$ and $V_h = k_h V_s$, where $k_l$ and $k_h$ are the feedforward gains. In this control scheme, the amplitude of the ramp, $V_h - V_l = (k_h - k_l)V_s$, is proportional to the source voltage.

### A. Prevention of Period Doubling Bifurcation

Define the following function of $d$:

$$H(d) =: 2\text{Re}[-\sum_{k=1}^{\infty} G(j(k-\frac{1}{2})\omega_s)+(1-e^{jk\omega_s d})G(jk\omega_s)] \tag{16}$$

Denote its maximum by $H_{\max}$ and its minimum by $H_{\min}$.

The condition for bifurcation in Eqn. (11) becomes

$$H(d) = k_h - k_l \tag{17}$$

If the values of $k_h$ and $k_l$ are chosen to satisfy $(k_h - k_l) > H_{\max}$ or $(k_h - k_l) < H_{\min}$, the period doubling bifurcation is prevented because the bifurcation condition is never met.

### B. Line Regulation

In the last subsection, one has two degrees of freedom in choosing the values of $k_h$ and $k_l$. This freedom can be used to achieve another objective besides preventing period doubling bifurcation. A common objective in DC-DC conversion is line regulation, which is addressed in this subsection.

First, an equation related to the duty cycle $D_c = 1 - d/T$ is derived. Assume the switching frequency is high enough that the high order terms in Eqn. (9) can be ignored because of the low-pass nature of $G(s)$. Then Eqn. (9) becomes

$$V_s = \frac{V_l + (V_h - V_l)\frac{d}{T} - gV_r}{(1-\frac{d}{T})G(0)} \tag{18}$$

Solving this equation for $d/T$ gives

$$\frac{d}{T} = \frac{G(0)V_s - V_l + gV_r}{G(0)V_s - V_l + V_h} \tag{19}$$

Next, the average output voltage, denoted as $[v_o]_{\text{AVE}}$, can be related to the source voltage $V_s$ as

$$\begin{aligned} [v_o]_{\text{AVE}} &= [v_{\text{d}}]_{\text{AVE}}, \text{ from Fig. 2 and } G_1(0) = 1 \\ &= (1-\frac{d}{T})V_s, \text{ from Eqn. (7)} \\ &= (\frac{V_h - gV_r}{G(0)V_s - V_l + V_h})V_s, \text{ from Eqn. (19)} \\ &= \frac{k_h V_s - gV_r}{G(0) - k_l + k_h} \end{aligned} \tag{20}$$

Two approaches make output voltage independent of source voltage: large $G(0)$ (such as integral control) or simply setting $k_h = 0$. The second approach is pursued here. From Eqn. (20) with $k_h = 0$, one has

$$k_l = G(0) + \frac{gV_r}{[v_o]_{\text{AVE}}} \tag{21}$$

Thus, given a buck converter power stage ($G_1(s)$) with desired average output $[v_o]_{\text{AVE}}$, and an error amplifier ($g$ and $G_2(s)$), line regulation can be achieved by adding a feedforward loop to adjust the ramp $h(t)$ by setting $V_h = 0$ and $V_l = k_l V_s$ with $k_l$ as in Eqn. (21). For the case in Example 1, $k_l = g_1(1 - V_r/[v_o]_{\text{AVE}})$.

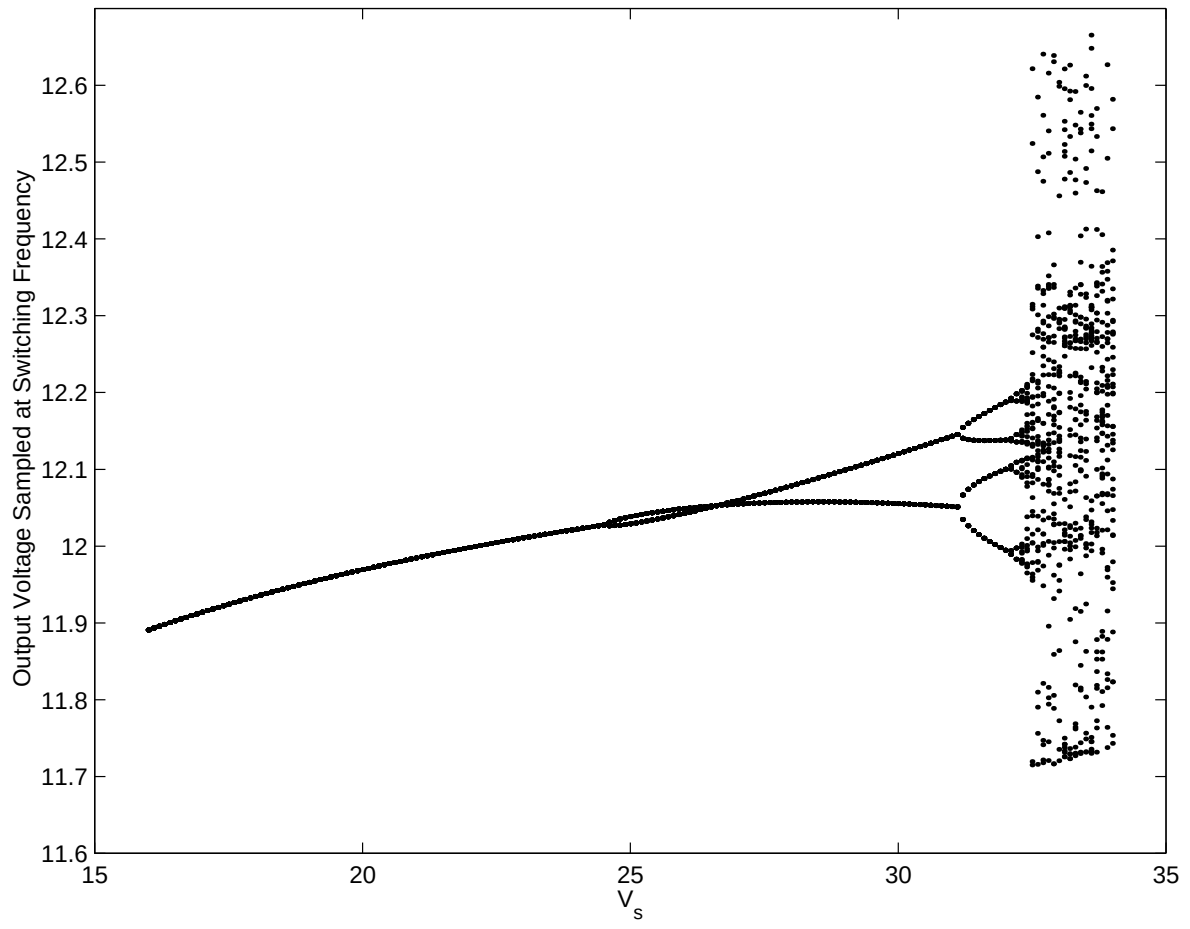

Figure 10: Bifurcation diagram of the circuit in Fig. 9

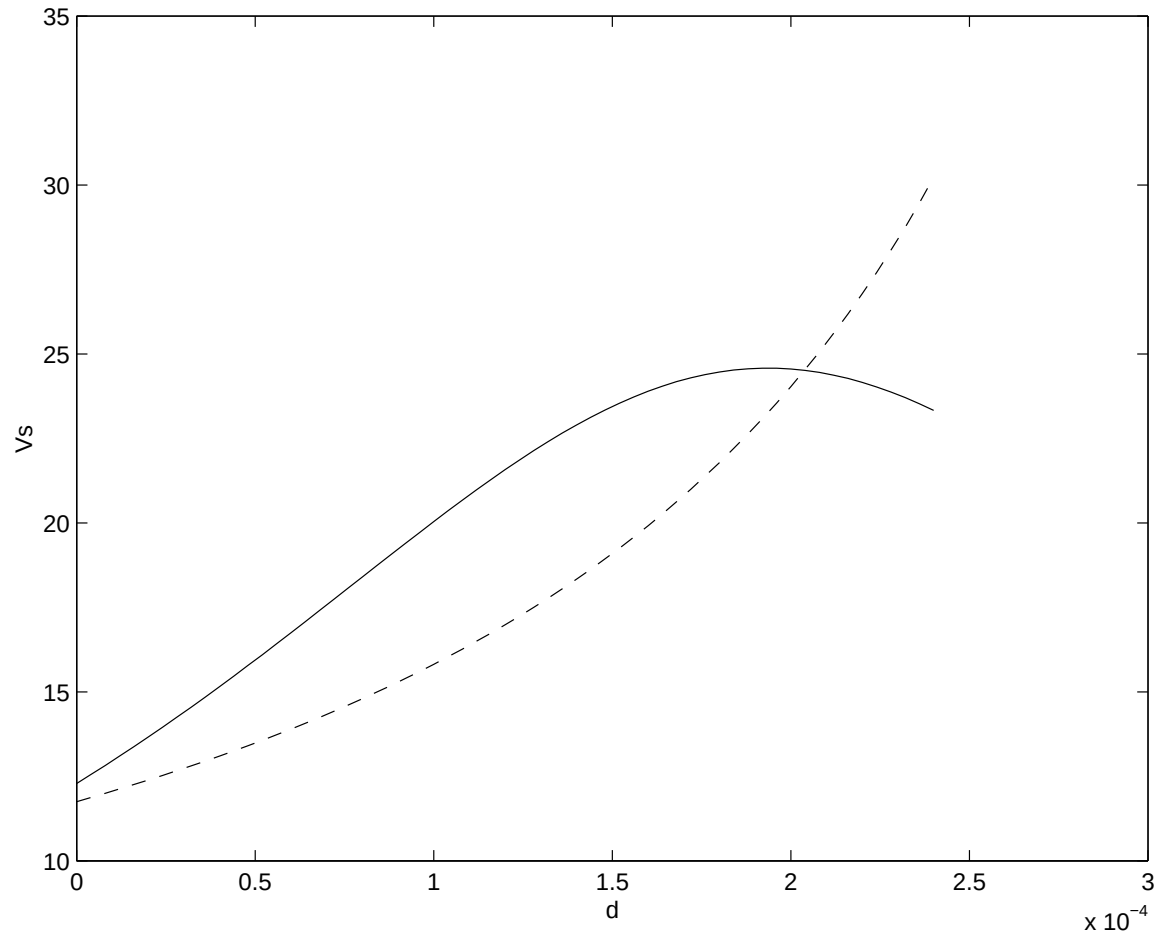

Figure 11: Dashed line: Eqn. (9); solid line: Eqn. (11)

## C. Combined Period Doubling Prevention and Line Regulation

If $k_h = 0$ and the value of $k_l$ given by Eqn. (21) satisfies $k_l < -H_{\max}$ or $k_l > -H_{\min}$, both prevention of period doubling bifurcation and line regulation are achieved.

## V. ILLUSTRATIVE EXAMPLE

Consider the buck converter in Example 1 with the same system parameters as (Hamill *et al.*, 1992): $T = 400$ $\mu$s, $L = 20$ mH, $C = 47$ $\mu$F, $R = 22$ $\Omega$, $V_r = 11.3$ V, $g_1 = 8.4$, $V_l = 3.8$ V, $V_h = 8.2$ V, (then $h(t) = 3.8 + 4.4[\frac{t}{T} \bmod 1]$). Let $V_s$ be the bifurcation parameter. Here the focus is on determination of the bifurcation point and controller design, although the switching frequency is low.

It has been shown that a period doubling bifurcation occurs at $V_s = 24.5$ through simulation (Hamill *et al.*, 1992) and through calculating the locus of the eigenvalues of a discrete-time model (Fossas and Olivar, 1996; Fang, 1997). The bifurcation diagram is shown in Fig. 10. From the diagram, the stable range of $V_s$ is short (16 V to 24.5 V). In this range of $V_s$, the output voltage varies from 11.9 V to 12.03 V as the source voltage varies.

The period doubling bifurcation point can be determined by plotting Eqns. (9) and (11) together (see Fig. 11). The intersection ($V_{s,*}$, $d_*$) =(24.5, $2.04 \times 10^{-4}$) of these graphs is the period doubling bifurcation point.

To show the effect of ESR $R_c$ and $T$ on the critical source voltage $V_{s,*}$, another two conditions ($R_c$=1 $\Omega$ and $T$=250 $\mu$s) are considered. The exact and estimated $V_{s,*}$ are shown in Table 1. It shows that larger $\omega_s$ leads to larger $V_{s,*}$ and ESR $R_c$ has little effect on $V_{s,*}$. It also shows that Eqn. (13), expressed explicitly in terms of system parameters, gives close estimate of $V_{s,*}$.

In the following, a feedforward scheme is designed and added to the original controlled buck converter to achieve two objectives: (i) a wider $V_s$ operating range (16 V to 35 V) without period doubling bifurcation and (ii) line regulation with $[v_o]_{\mathrm{AVE}} = 10$ V.

Table 1: Determination of $V_{s,*}$ for different $T$ and $R_c$

| | | | |
|---|---|---|---|
| Switching period $T$ ($\mu$s) | 400 | 400 | 250 |
| ESR $R_c$ ($\Omega$) | 0 | 1 | 0 |
| Exact $V_{s,*}$ by Eqns. (9), (11) | 24.5 | 24.9 | 49.5 |
| Estimate of $V_{s,*}$ by Eqn. (12) | 20.2 | 22.4 | 51.8 |
| Estimate of $V_{s,*}$ by Eqn. (13) | 20.2 | 21.2 | 51.8 |

A plot of $H(d)$ is shown in Fig. 12 (here, $H_{\max} = 0.358$ and $H_{\min} = 0.1792$). Set $k_h = 0$. From Eqn. (21), $k_l = -1.092$. The condition $k_l < -H_{\max}$ is satisfied. Thus both prevention of period doubling bifurcation and line regulation can be achieved.

A feedforward control to adjust the ramp amplitude is designed. The ramp signal has $V_l = 0$ and $V_h = -1.092V_s$. The resulting bifurcation diagram of the buck converter with the feedforward control is shown in Fig. 13. Compared with Fig. 10, Fig. 13 shows a wider operating range for $V_s$ (16 V to 35 V) and good line regulation, with $[v_o]_{\mathrm{AVE}} = 10$ V.

Take $V_s = 28$ V for example. The output voltage response during start-up (starting from $(i_L, v_o) = (0, 0)$) is shown in Fig. 14. The output voltage is regulated to 10 V. The switching operation depends on intersection of the signals $h(t)$ and $y(t)$. These are shown in Fig. 15. Different from the traditional positive ramp, the signal $h(t)$ is negative. It is because the regulated output voltage is smaller than $V_r$, and a negative $h(t)$ is needed to compare with negative $y = g_1(v_o - V_r)$.

## VI. CONCLUSIONS

Exact harmonic balance analysis is applied to study period doubling bifurcation in the buck converter in continuous conduction mode. A simple and unified dynamic model of the buck converter under voltage mode or current mode control is derived. The model is general that it can be applied to the buck converter with various configurations, such as with a second output filter, with ESR $R_c$, or with a high-order error amplifier. This model consists of the feedback connection of a linear system and a nonlinear one. An exact condition for period doubling bifurcation is given in terms of solving a pair of algebraic equations. The critical condition can be approximated explicitly in terms of system parameters, as in Eqns. (13)-(15). These conditions for buck converters show that larger values of switching frequency and ramp amplitude lead to larger stable range of source voltage. ESR $R_c$ does not affect stability in voltage mode control, but it does in current mode control. To prevent the period doubling bifurcation, a ramp-adjusting feedforward control is designed. A wider operating range of source voltage is achieved, along with line regulation. Simulations are given to illustrate the effectiveness of the design technique.

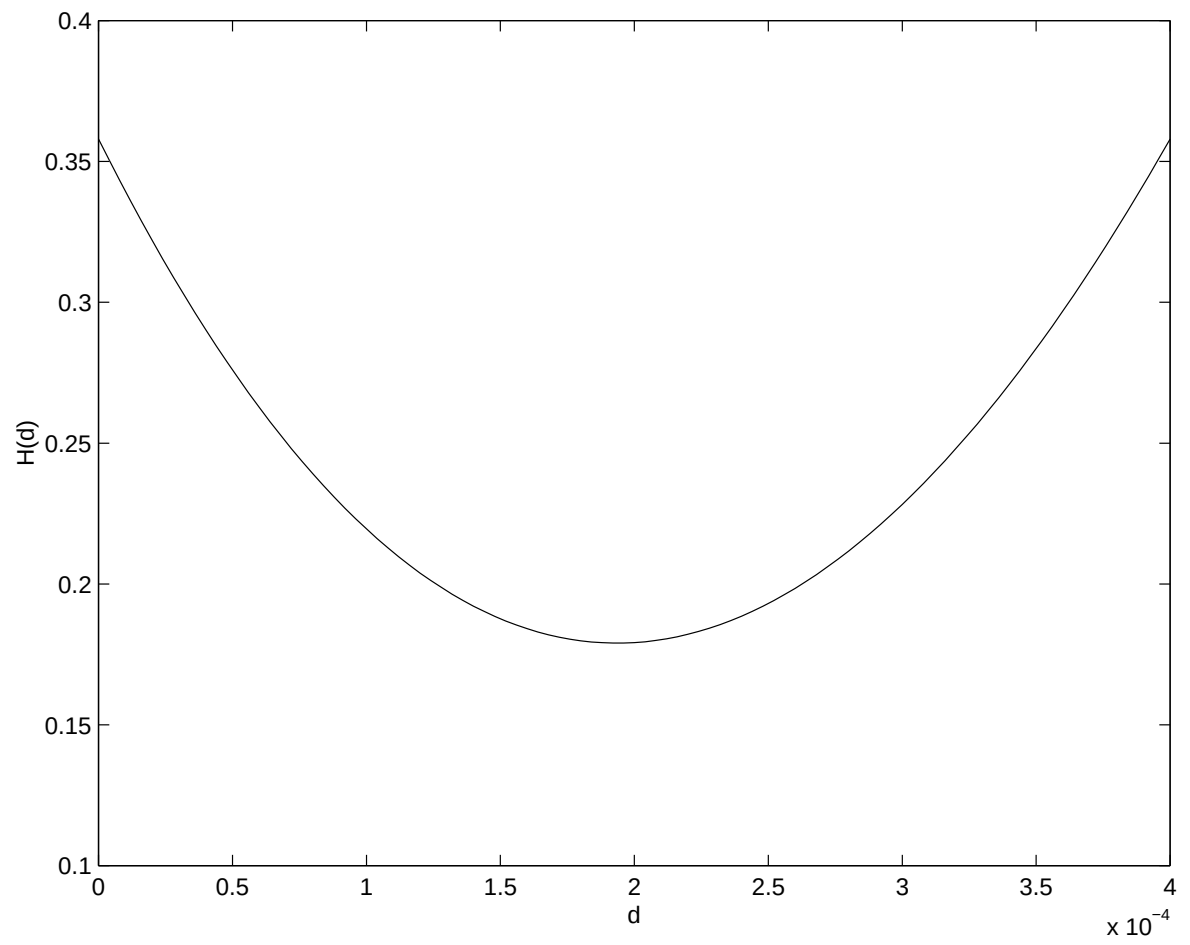

Figure 12: Plot of function $H(d)$

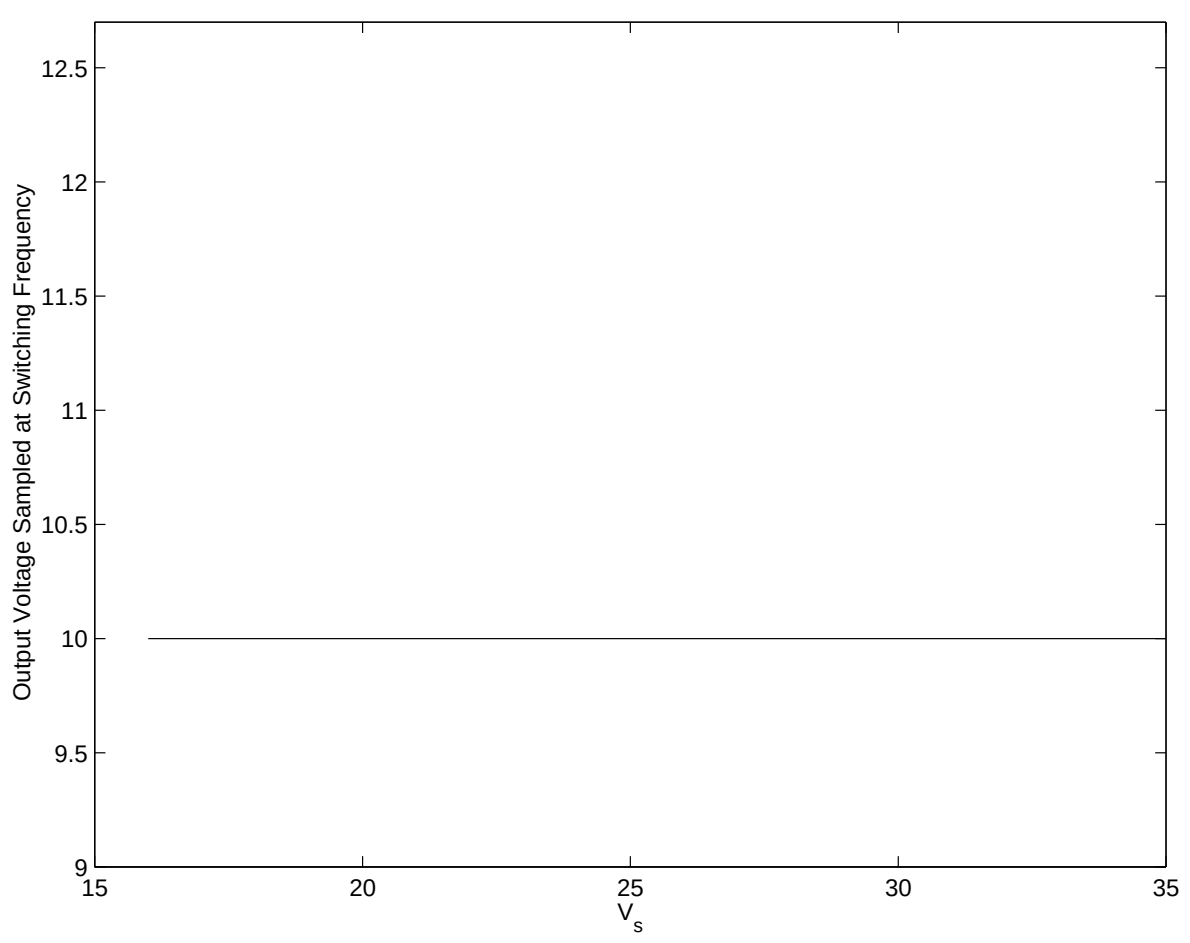

Figure 13: Bifurcation diagram (feedforward control)

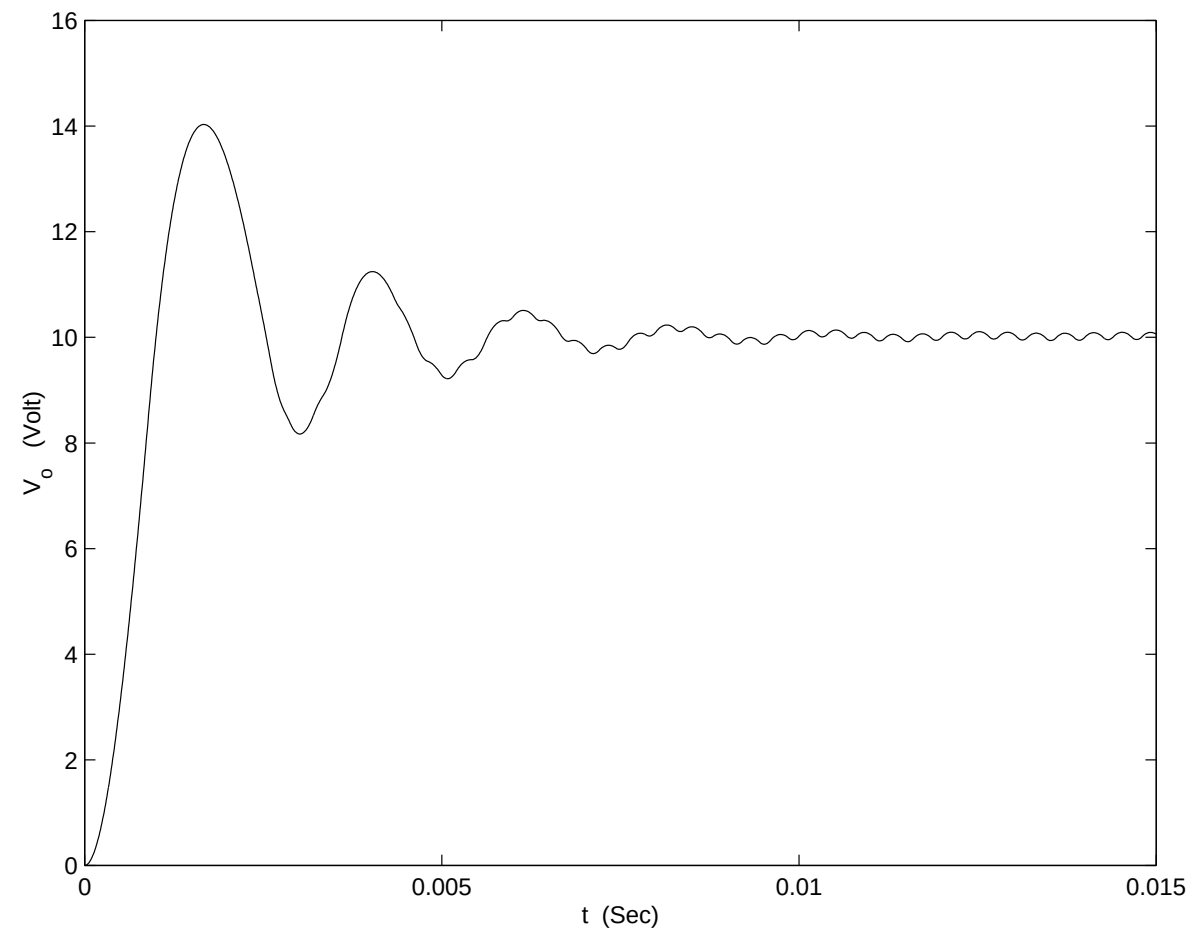

Figure 14: Output response with feedforward control

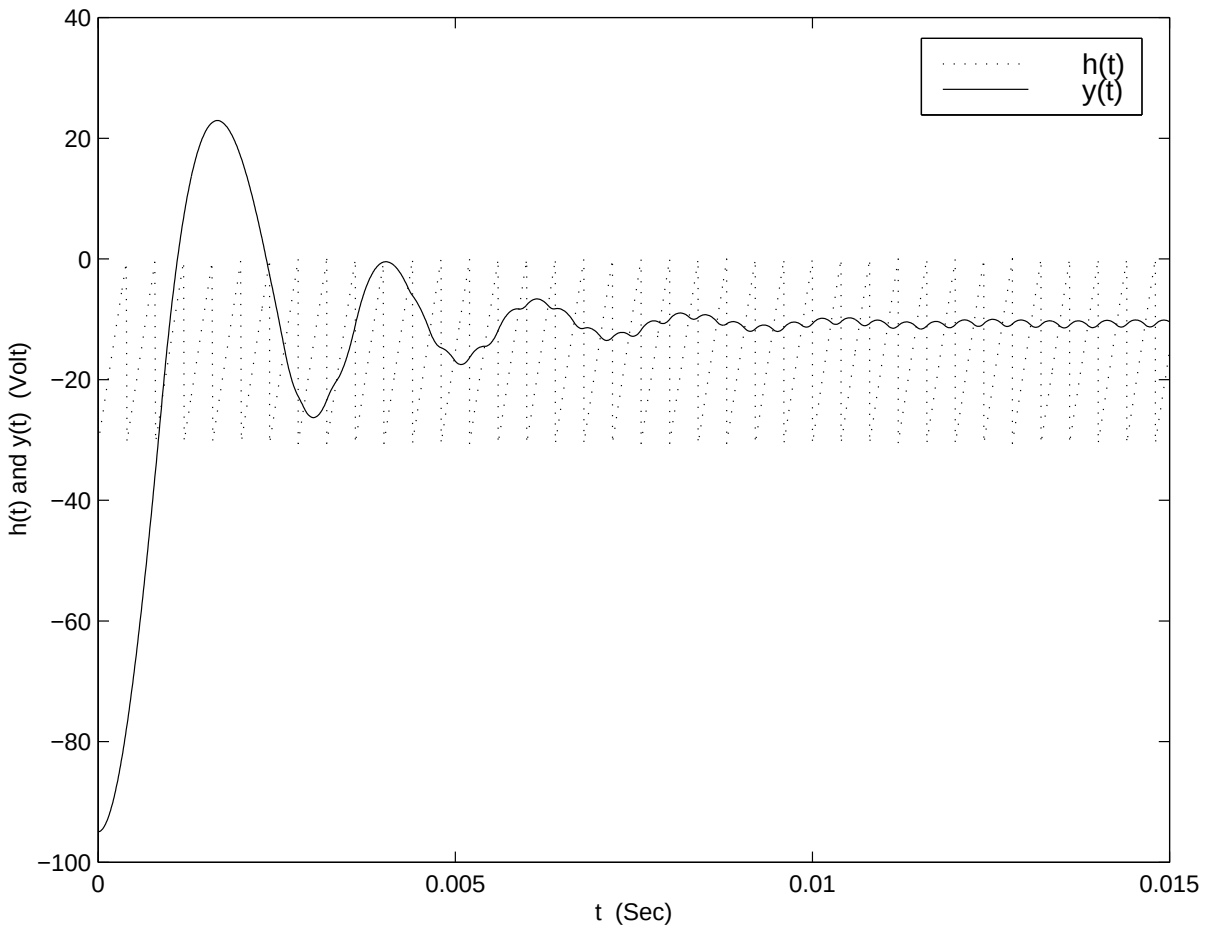

Figure 15: $h(t)$ and $y(t)$ with feedforward control

## ACKNOWLEDGMENTS

This research has been supported in part by the Office of Naval Research under Multidisciplinary University Research Initiative (MURI) Grant N00014-96-1-1123, and by the U.S. Air Force Office of Scientific Research under Grant F49620-96-1-0161.